\documentclass[%
 reprint,
 amsmath,amssymb,
 aps,
prl
]{revtex4-1}

\usepackage{graphicx}
\usepackage{dcolumn}
\usepackage{bm}

\usepackage[frozencache=true,cachedir=minted-cache]{minted}

\begin{document}


\title{Quantum Elliptic Vortex in a Nematic-Spin Bose-Einstein Condensate
 }

\author{Hiromitsu Takeuchi}
\email{takeuchi@osaka-cu.ac.jp}
\homepage{http://hiromitsu-takeuchi.appspot.com}
\affiliation{
Department of Physics and Nambu Yoichiro Institute of Theoretical and Experimental Physics (NITEP),\\
 Osaka City University, Osaka 558-8585, Japan
}




\date{\today}

\begin{abstract}
  We find a novel topological defect in a spin-nematic superfluid theoretically.
A quantized vortex spontaneously breaks its axisymmetry, leading to an elliptic vortex in nematic-spin Bose-Einstein condensates with small positive quadratic Zeeman effect.
 The new vortex is considered the Joukowski transform of a conventional vortex.
Its oblateness grows when the Zeeman length exceeds the spin healing length.
This structure is sustained by balancing the hydrodynamic potential and the elasticity of a soliton connecting two spin spots,
 which are observable by {\it in situ} magnetization imaging.
The theoretical analysis clearly defines the difference between half quantum vortices of the polar and antiferromagnetic phases in spin-1 condensates.

\end{abstract}

\pacs{Valid PACS appear here}
\maketitle

Topological defects (TDs) caused by spontaneous symmetry breaking (SSB) phase transition is ubiquitous,
 existing as skyrmions in spintronic devices \cite{wiesendanger2016nanoscale},
  vortices in superconductors and superfluids \cite{tilley2019superfluidity,donnelly1991quantized},
   and even disclinations in LCD displays \cite{chandrasekhar_1992}.
Thanks to the universal concept of SSB,
TDs in laboratories are useful for simulating TDs in other exotic settings, the early universe, the dense matters in compact stars, and higher-dimensional spacetimes in field theory \cite{vilenkin2000cosmic, page2006dense,vachaspati2006kinks,doi:10.1142/S0217751X0502519X}.
Multicomponent superfluids with spin freedom, such as spin-triplet superfluid $^3$He and binary and spinor Bose-Einstein condensates (BECs) \cite{vollhardt2013superfluid,volovik2003universe, kasamatsu2005vortices, kawaguchi2012spinor},
are powerful tools to develop theories of TDs since various TDs are realized there.
Such superfluids are called the nematic-spin superfluids \cite{nematic-spin},
 whose order state is partly represented by a vector $\hat{\bm d}$ that mimics the {\it director} $\tilde{\bm d}$ in nematic liquid crystals (NLCs) \cite{chandrasekhar_1992}.

Nematic-spin superfluids support not only conventional TDs in NLCs
(disclination, hedgehog, domain wall, and boojum \cite{merminRevModPhys.51.591,kumar1989certain,volovik1992exotic,mineyevPhysRevB.18.3197, zhouPhysRevLett.87.080401, zhou2003quantum, ruostekoskiPhysRevLett.91.190402, Mermin1977, volovik1990defects, misirpashaev1991topological, takeuchi_doi:10.1143/JPSJ.75.063601}),
 but also novel TDs combined with the superfluidity, e.g., half quantum vortex (HQV) \cite{Leonhardt:2000km}.
The term HQV is used also in exciton-polariton condensates \cite{PhysRevLett.99.106401,Lagoudakis974}.
The simplest type of HQV has been realized experimentally in different superfluids \cite{matthewsPhysRevLett.83.2498,seoPhysRevLett.115.015301,auttiPhysRevLett.117.255301},
where the core of a vortex in a spin component is occupied by other components.
A nontrivial HQV is terminated by a domain wall across which the order-parameter phase jumps by $\pi$.
The wall-HQV composites were first realized as the double-core vortices in $^3$He-B \cite{kondoPhysRevLett.67.81} and revisited \cite{makinen2019half,volovikPhysRevResearch.2.023263,zhang2020one}, motivated by the early Universe scenario nucleating the composites of the Kibble-Lazarides-Shafi (KLS) walls and cosmic strings \cite{kibblePhysRevD.26.435,KIBBLE1982237,KIBBLE1982237,bunkov2000topological}.
Recently, the nonequilibrium dynamics of wall-HQV composites were observed in phase transition from the antiferromagnetic (AF) phase to the polar (P) phase \cite{PhaseName} in a spin-1 $^{23}$Na BEC \cite{kangPhysRevLett.122.095301,kangPhysRevA.101.023613}.
However, the dynamics are poorly understood, because of the lack knowledge about properties of wall-HQV composites.
 Determining these properties is important for understanding KLS-wall-HQV composites and double-core vortices in $^3$He-B \cite{PhysRevB.66.224515,PhysRevLett.115.235301,PhysRevB.99.104513,PhysRevB.101.094512,PhysRevB.101.024517}), the Berezinskii-Kosterlitz-Thouless transition in spinor BECs \cite{mukerjeePhysRevLett.97.120406,jamesPhysRevLett.106.140402,kobayashi_doi:10.7566/JPSJ.88.094001},
 and even the quark-confinement problem in hadronic physics connected with the vortex-confinement problem \cite{sonPhysRevA.65.063621,tylutkiPhysRevA.93.043623,etoPhysRevA.97.023613,GallemiPhysRevA.100.023607}.

\begin{figure}
\begin{center}
\includegraphics[width=1.0 \linewidth, keepaspectratio]{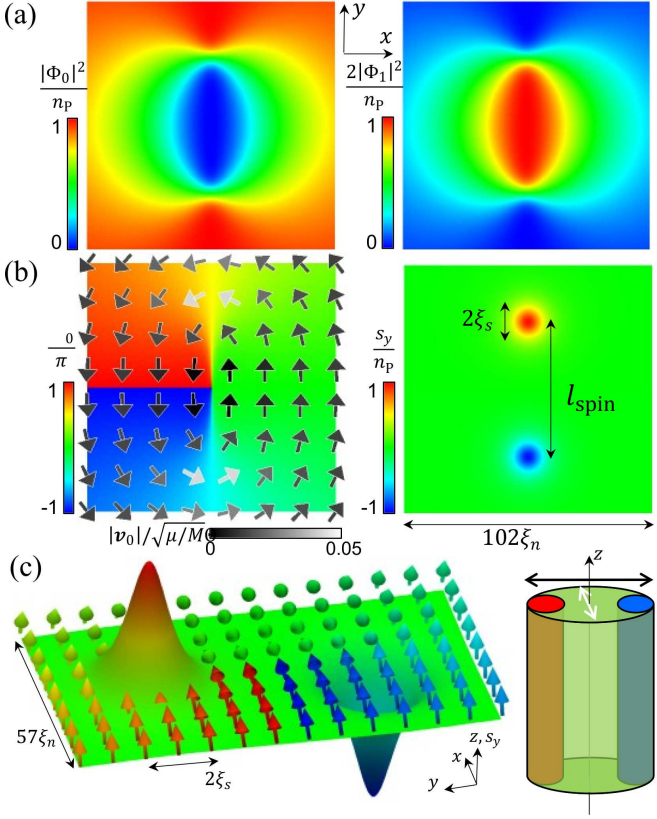}
\end{center}
\vspace{-5mm}
\caption{
The cross-sectional profile of an elliptic vortex for $q/\mu=2^{-17}\approx 7.6\times 10^{-6}$.
(a) The left and right sides show the profiles of $\frac{|\Phi_0|^2}{n_{\rm P}}$ and $\frac{|\Phi_1|^2}{2n_{\rm P}}$, respectively.
The density $|\Phi_{-1}|^2$ (not shown) is the same as $|\Phi_{1}|^2$.
(b) The vector field on the left shows ${\bm v}_0=\frac{\hbar}{M}{\bm \nabla}\Theta_0$,
 with the background plot of $\Theta_0 =\arg \Phi_0$.
The phase $\arg \Phi_{\pm 1}$ is homogeneous inside the core (not shown).
The spin density $s_y$ is plotted on the right, while $s_x=s_z=0$.
(c) Left: the texture of the unit vector ${\bm g}/|{\bm g}|$ (arrow) in Eq.~(\ref{eq:dtexture}).
The color of the arrows and the surface correspond to $\Theta_0$ and $s_y$ in (b), respectively.
Right: schematic of the three-dimensional structure of the vortex core.
 The distance $l_{\rm spin}$ between the two spin spots with opposite transverse magnetization (blue and red) is determined by the balance between the hydrodynamic potential and the elastic potential by the AF soliton (green).
 The width ($\sim l_{\rm spin}$) and thickness of the core are represented by black and white arrows, respectively.
}
\label{Fig_dv}
\end{figure}

Here, it is theoretically shown that
a wall-HQV composite in spin-1 BECs \cite{kangPhysRevLett.122.095301,kangPhysRevA.101.023613} takes an exotic state in equilibrium with a small positive quadratic Zeeman effect.
This state, called the elliptic vortex, is hydrodynamically considered the Joukowski transform of a conventional vortex and has an elliptic structure with spin spots (Fig.~\ref{Fig_dv}).
The spots are confined to the elliptic-vortex core and stabilized by a balance between the hydrodynamic effect and the tension of a domain wall or a soliton spanned between the spots.

{\it Formulation}.---A spin-1 BEC is described by the condensate wave function $\Phi_m~(m=0,\pm 1)$ of the $| m \rangle$ Zeeman component in the Gross-Pitaevskii model \cite{kawaguchi2012spinor,pethick2008bose}.
The thermodynamic energy is represented as
$
G(\{ \Phi_m \})=
\int d^3x {\cal G}$, with
${\cal G}= \frac{\hbar^2}{2M}\sum_m|{\bm \nabla}\Phi_m|^2+ {\cal U}$,
and
\begin{eqnarray}
{\cal U}=\frac{c_0}{2}n^2
+\frac{c_2}{2}{\bm s}^2
 -(\mu-q) n-q\left| \Phi_0 \right|^2-ps_z.
\label{eq:energy_density}
\end{eqnarray}
Here, we introduced the chemical potential $\mu (>0)$ and the coefficient $q$ ($p$) of the quadratic (linear) Zeeman effect.
In the Cartesian representation  ${\bm \Phi}=\left[\Phi_x,\Phi_y,\Phi_z\right]^{\rm T}=\left[\frac{-1}{\sqrt{2}}(\Phi_{+1}-\Phi_{-1}),\frac{-i}{\sqrt{2}}(\Phi_{+1}+\Phi_{-1}),\Phi_0 \right]^{\rm T}$ \cite{ohmi1998bose},
the condensate density is expressed by the dot product $n=\sum_m \left| \Phi_m \right|^2={\bm \Phi}^*\cdot{\bm \Phi}$ and the spin density by the cross product ${\bm s}=\left[s_x,s_y,s_z \right]^{\rm T}=i {\bm \Phi}\times {\bm \Phi}^*$.

The ground (bulk) state is obtained by minimizing $U=\int d^3x {\cal U}$.
Assuming $c_2=0.016c_0>0$ with $p=0$ obtained experimentally \cite{kangPhysRevLett.122.095301,kangPhysRevA.101.023613},
the ground state is in the P state
$
   {\bm \Phi}={\bm \Phi}_{\rm P}=\left[0,0, \sqrt{n_{\rm P}}e^{i\theta_{\rm G}}\right]^{\rm T}
   \label{eq:Pstate}
$
with the bulk density $n_{\rm P}=\frac{\mu}{c_0}$ and the order-parameter phase $\theta_{\rm G}$.
By rescaling energy and length by $\mu$ and $\xi_n\equiv \frac{\hbar}{\sqrt{M\mu}}$, respectively,
 the P phase is parametrized by two dimensionless quantities $\frac{c_2}{c_0}$ and $\frac{q}{\mu}$.

 {\it Vortex core structure}.---One might expect that there is nothing strange about the occurrence of vortices in P phase,
 whose order parameter (OP) is a complex scalar $\Phi_0(=\Phi_z)$ with $\Phi_{\pm 1}=0$,
  as in conventional superfluids.
However, the core of a singly quantized vortex can be unconventional in multicomponent superfluids,
 occupied by other components so as to reduce the condensation energy, e.g., $^3$He-B at high pressure \cite{PhysRevLett.51.2040} and segregated binary BECs \cite{PhysRevA.87.063628}.
Similarly, the vortex core can be occupied by the $m=\pm 1$ component in the P phase.

\begin{figure}
\begin{center}
\includegraphics[width=1.0 \linewidth,keepaspectratio]{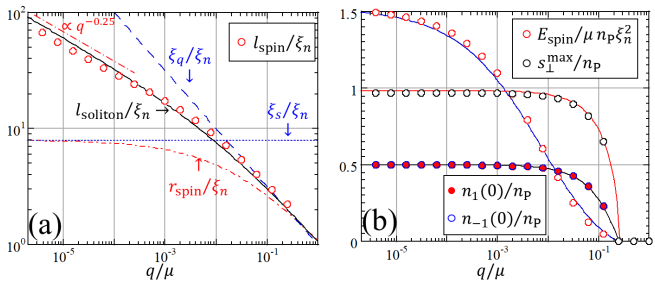}
\end{center}
\vspace{-5mm}
\caption{
(a) The $q$ dependence of $l_{\rm spin}$.
The solid curve represents the evaluation by Eq.~(\ref{eq:E_soliton}).
All lengths are rescaled by $\xi_n$.
(b) The $q$ dependence of the spin interaction $E_{\rm spin}$, the maximum spin density $s_\bot^{\rm max}=\max(s_y)$, and the core density $n_{\pm 1}(0)$.
The solid curve tracing the data corresponds to an analytic formula (see text).
}
\label{Fig_dtexture}
\end{figure}

To examine the conjecture,
the lowest-energy solution was obtained by numerically minimizing $G$ in the steepest descent method \cite{curry1944method}.
It is found that a nonaxisymmetric core structure is observed for small $q/\mu$.
Figure~\ref{Fig_dtexture} shows the typical cross-sectional profile of the vortex for $\frac{q}{\mu}=2^{-17}$ in a cylindrical flat-bottom potential of sufficiently large radius \cite{NumMethod}.
 The vortex core is occupied by the $m=\pm 1$ components, and the density $n$ is mostly homogeneous [Fig.~\ref{Fig_dv}(a)].
Surprisingly, the velocity field forms an elliptic structure, and two spin spots are observed with opposite transverse magnetization ($s_y \neq 0$) at the edges of the core [Fig.~\ref{Fig_dv}(b)].
Since the order-parameter phase $\Theta_0(=\arg \Phi_0)$ jumps by $\pi$ across the $x=0$ plane and rotates by $\pi$ around each spin spot, this structure is regarded as a wall-HQV composite composed of a wall and two HQVs with the same circulation.

The distance $l_{\rm spin}$ between the spin spots is a decreasing function of $q$ [Fig.~\ref{Fig_dtexture}(a)].
Accordingly, the density $n_{\rm core}$ at the center of the vortex core and the maximum spin density $s_\bot^{\rm max}$ decrease with $q$ and vanish at a critical value $q_{\rm C}\approx 0.25 \mu$ [Fig.~\ref{Fig_dtexture}(b)]
\cite{Qc_vortex,underwood2020properties}.
 This behavior is similar to that of the AF-core soliton \cite{liu2020phase},
 where the soliton core is vacant for large $q$ but occupied by the local AF state (${\bm s}=0$ with $\Phi_{\pm 1}\neq 0$ and $\Phi_0\approx 0$) for small $q$.
In our case, however, the vortex core is occupied by two different sates, the local broken-axisymmetry (BA) state (${\bm s}\bot \hat{\bm z}$ with $\Phi_{1}\Phi_0\Phi_{-1}\neq 0$) and the local AF state.

In order to explain the nematic-spin order in the vortex core,
we extend the OP space as
\begin{eqnarray}
  {\bm \Phi}=\sqrt{n}e^{i\theta_{\rm G}}\hat{\bm d},
  \label{eq:director}
\end{eqnarray}
which represents the OP in the ground state with ${\bm s}=0$ for $q=0$.
The real unit vector $\hat{\bm d}$ is called the {\it pseudo}-director;
 the state of $(\hat{\bm d},\theta_{\rm G})$ is identical to $(-\hat{\bm d},\theta_{\rm G}+\pi)$.
In terms of the extended OP, the ground state in the P (AF) phase with $q>0$ ($ q < 0 $) is represented as
   $n=n_{\rm P}$ and $\hat{\bm d}=\pm \hat{\bm z}$ ($n=n_{\rm AF}$ and $\hat{\bm d}=\hat{\bm r}_\bot$) within the unit vector $\hat{\bm z}$ ($\hat{\bm r}_\bot$) parallel (normal) to the quantization axis and the density $n_{\rm AF}=\frac{\mu-q}{c_0}$ of the AF state.
To describe the magnetization together with the nematic-spin order,
it is useful to introduce a representation
\begin{eqnarray}
  {\bm \Phi}=e^{i\Theta_0}({\bm g}+i {\bm h})
\label{eq:dtexture}
\end{eqnarray}
 with ${\bm g}=[g_x,g_y,g_z]^{\rm T}$ with $g_z\geq 0$ and ${\bm h} \bot \hat{\bm z}$.
Equation (\ref{eq:dtexture}) reduces Eq.~(\ref{eq:director}) for ${\bm s}=2{\bm g}\times {\bm h}=0$ with ${\bm g}=\sqrt{n}\hat{\bm d}$. 
The left panel in Fig.~\ref{Fig_dv}(c) shows a cross-sectional plot of $\frac{{\bm g}}{|{\bm g}|}$ and $s_y$.
In the region between the spin spots, ${\bm g}$ lies on the $xy$ plane forming the local AF state,
where the state $(\hat{\bm d},\Theta_0)=(-\hat{\bm x},\pm\pi)$ for $x > 0$ is identical to $(\hat{\bm x},0)$ for $x < 0$ along the $yz$ plane.
The nematic-spin order is destroyed when $\hat{\bm d}$ is ill-defined in the spin spots occupied by the local BA state [see the right panel in Fig.~\ref{Fig_dv}(c)].

To clarify our problem,
the main goal is to answer the following two questions:
\begin{description}
  \item[Q1] What causes the axisymmetry breaking?
  \item[Q2] What is the physical mechanism to stabilize the elliptic structure?
\end{description}

{\it Vortex winding rule}.---As the answer for the first question, it is claimed that the spin interaction breaks the axisymmetry.
To justify the claim logically, we introduce a winding rule of an axisymmetric vortex in spin-1 BECs.
We consider a straight vortex along the $z$ axis, the cross section of which is axisymmetric as the ansatz $\Phi_m=f_m(r)e^{iL_m\varphi}$,
 with radius $r=\sqrt{x^2+y^2}$, and azimuthal angle $\varphi$ in cylindrical coordinates.
The rule states that $L_m$ is parametrized by the winding numbers $L$ and $N$, associated with the mass and spin current, respectively, and given by
\begin{eqnarray}
  L_m=L+mN~~~~~(L,N=0,\pm 1, \pm 2,...).
\label{eq:winding}
\end{eqnarray}
The rule is related to the phase factor $\delta\Theta=(L_{+1}+L_{-1}-2L_{0})\varphi$. By substituting the ansatz into the equation of motion, we have, for the equation of $\Phi_0$, $0=(h_0-\mu+c_0n+c_2f_{+1}^2+c_2f_{-1}^2+2c_2f_{+1}f_{-1}e^{i\delta\Theta})f_0$.
The last term comes from the transverse spin density and
the equation of real function $f_m$ is solved when $e^{i\delta\Theta}=\pm 1$, resulting in Eq.~(\ref{eq:winding}).
Therefore, this rule is applicable for $s_x\neq 0$ or $s_y\neq 0$ with $f_{+1}f_{-1}f_0\neq 0$ \cite{Winding_rule}.

By contraposition of the above argument,
the vortex must be nonaxisymmetric,
 when the winding rule is not satisfied.
As seen in Fig.~\ref{Fig_dv}, only the $m=0$ component has a nonzero winding number,
 corresponding to $L_0=1$ and $L_{\pm 1}=0$.
Such a set of winding numbers cannot satisfy the winding rule.
The axisymmetry is exactly recovered only for $\Phi_{\pm 1}=0$ ($q \geq q_{\rm C}$).
Since the winding rule works for $s_x\neq 0$ or $s_y\neq 0$,
 the transverse magnetization appear as a manifestation of the axisymmetry breaking.
 The orientations of the transverse spin and the axes of the elliptic structure depend on the phases $\arg\Phi_m$.

{\it Joukowski mapping}.---To answer the second question, the potential flow theory in two-dimensional flow is extended to our problem.
The elliptic core structure hints at the Joukowski transformation \cite{milne1973theoretical},
since the velocity field on the cross section is considered a two-dimensional potential flow.
 This perception is the motivation for investigating the problem,
 and the following analysis leads to a quantitative evaluation of the core structure.

The velocity field ${\bm v}_0=\frac{\hbar}{M}{\bm \nabla}\Theta_0=(u,v)$ in the $xy$ plane is generated by a conformal mapping
 called the Joukowski transformation from a vortex within a cylinder of radius $a$ in the $\zeta$ complex plane to the $xy$ plane,
$x+iy=\zeta+\frac{a^2}{\zeta}$ \cite{milne1973theoretical}.
By using the parametrization $\zeta=ia e^{\phi+i\psi}~(\phi \geq 0)$, one obtains $(x,y)=2a(\cosh\phi \cos(\psi+\pi),\sinh\phi \sin(\psi+\pi) )$,
 representing an ellipse of {\it width} $4a\cosh\phi$ and {\it thickness} $4a\sinh\phi$. 

The velocity field is computed by applying the conformal mapping to the complex velocity potential $W$ of the vortex in the $\zeta$ plane,
\begin{eqnarray}
  W=-i\frac{\kappa}{2\pi}\log \zeta.
\label{eq:velocity}
\end{eqnarray}
The circulation $\kappa=\frac{2\pi\hbar}{M}$ around a quantized vortex is conserved in the transformation as follows.
By applying the transformation to Eq.~(\ref{eq:velocity})
 and using  the formula
 $v \to \pm \frac{\kappa}{2\pi}\frac{1}{\sqrt{4a^2-y^2}}$ for $|y|<2a$ in the limit $x\to \pm 0$,
  the vorticity $\omega_z(x,y)=({\bm \nabla}\times {\bm v}_0)_z$ forms a segment singularity of width $4a$,
\begin{eqnarray}
\omega_z({\bm r})=\frac{\kappa}{\pi} \frac{1}{\sqrt{4a^2-y^2}} \delta(x) \Theta(2a-|y|).
\label{eq:vorticity}
\end{eqnarray}
with the step function $\Theta$ ($\Theta=1$ for $2a \geq |y|$ and $\Theta=0$ for $2a < |y|$).
By integrating Eq.~(\ref{eq:vorticity}), it is confirmed that the circulation is conserved as $\int dxdy\omega_z=\kappa$.

{\it Hydrodynamic potential}.---To reveal the physical mechanism that stabilizes the elliptic vortex,
 the energy $E_{\rm vortex}$ of a vortex of unit length is evaluated.
The vortex energy in the $\zeta$ plane is computed conventionally by considering the contribution from the core region ($|\zeta|<\rho_{\rm core}\equiv ae^{\phi_{\rm core}}$) and the outer region ($|\zeta| > \rho_{\rm core}$) separately \cite{donnelly1991quantized}.
Similarly, we consider the Joukowski mapping of the former and the latter, corresponding to an ellipse of area $S_{\rm core}$ and outer area $S_{\rm out}$ in the $xy$ plane, respectively.

The core region is characterized by two parameters $a$ and $r_{\rm core}\equiv \rho_{\rm core}-a$ as
\begin{eqnarray}
S_{\rm core}=\pi R_+ R_-=\pi\frac{(a+r_{\rm core})^4-a^4}{(a+r_{\rm core})^2},
\label{eq:S_core}
\end{eqnarray}
with $R_\pm=\frac{( a+r_{\rm core} )^{2} \pm a^2}{a+r_{\rm core}}$.
Here, $2R_{+(-)}$ is the width (thickness) of the ellipse.
For high oblateness with $\frac{a}{r_{\rm core}} \gg 1$,
 we have $R_+ \approx 2a$ and $R_- \approx 2r_{\rm core}$.
 The axisymmetric limit $\frac{a}{r_{\rm core}}\to 0$ results in $R_+=R_- \to r_{\rm core}$.

The vortex energy is defined as the excess energy in the presence of the vortex,
 with respect to the bulk energy $E_{\rm bulk}={\cal U}_{\rm P}(S_{\rm in}+S_{\rm out})$ with energy density ${\cal U}_{\rm P}=-\frac{1}{2}\mu n_{\rm P}$ in the bulk P phase.
The vortex energy is then represented formally by
\begin{eqnarray}
E_{\rm vortex}=E_{\rm out}+E_{\rm core}-E_{\rm bulk}=U_{\rm core}+U_{\rm out}
\end{eqnarray}
with $E_{\rm core(out)}=\int_{\rm S_{\rm core(out)}}dxdy {\cal G}$ and $U_{\rm core(out)}=E_{\rm core(out)}-{\cal U}_{\rm P}S_{\rm core(out)}$.
The potential $U_{\rm out}$ of the outer region is evaluated by computing the integral in $E_{\rm out}$ analytically with an approximation $n\approx n_{\rm P}\left(1-\frac{M}{2\mu}{\bm v}_0^2\right)$, where the quantum pressure is neglected.
In the approximation up to the order of ${\cal O}\left(\frac{M}{2\mu}{\bm v}_0^2\right)$,
a straightforward computation yields
\begin{eqnarray}
U_{\rm out}\approx U_{\rm hyd}=\frac{Mn_{\rm P}\kappa^2}{4\pi}\ln \frac{R}{a+r_{\rm core}}.
\label{eq:U_hyd}
\end{eqnarray}
Here, we used the radius $R=a e^{\phi_{\rm out}}$ of the system boundary by assuming $R\gg a$ \cite{Computation_I}.

{\it Elastic core potential}.---The core potential $U_{\rm core}$ is determined by introducing a phenomenological model,
 where a soliton is spanned between the spin spots.
This model is justified by the fact that the phase gradient is mainly concentrated around the spin spots, consistent with the vorticity distribution (\ref{eq:vorticity}) [see also Fig.~\ref{Fig_dv}(b)];
 thus, the core structure between the spots is similar to that of the AF-core soliton \cite{liu2020phase}.
Accordingly, we write
\begin{eqnarray}
U_{\rm core}=E_{\rm soliton}+E_{\rm spin}
\label{eq:U_hyd}
\end{eqnarray}
where the soliton energy $E_{\rm soliton}$ is a function of the soliton length $l_{\rm soliton}\sim l_{\rm spot}$ and the spin interaction $E_{\rm spin}$ comes from the second term of Eq.~(\ref{eq:energy_density}).

The spin interaction is determined independently from the hydrodynamic argument, and thus $U_{\rm core}$ depends explicitly on $l_{\rm spot}$ through $E_{\rm soliton}$.
The size $r_{\rm spin}$ and the magnitude $s_\bot^{\rm max}=\max(s_y)$ of the spin spot are asymptotic to $\xi_s=\frac{\hbar}{\sqrt{ Mc_2 n_{\rm P} } }$ and $n_{\rm P}$, respectively, for $\xi_q\gg \xi_s \gg \xi_n$.
For $\xi_s \gg \xi_q \gtrsim \xi_n$,
the core density grows as $n_{\pm 1} (0) \propto 1-\frac{q}{q_{\rm C}}$ in the continuous phase transition \cite{PT_core},
 and the size $r_{\rm spin}$ must be bounded below the vortex core size $\lesssim \xi_q$.
Therefore, the size of a spin spot is simply parametrized as
\begin{eqnarray}
  r_{\rm spin}^{-1}=\xi_s^{-1}+C_{\rm spin}\xi_q^{-1}
\label{eq:r_spin}
\end{eqnarray}
with $C_{\rm spin}\sim {\cal O}(1)$.
In fact, the spin interaction, estimated by $E_{\rm spin}= \frac{1}{2}c_2 (s_y^{\rm max})^2 \pi r_{\rm spin}^2$ agrees well with the numerical result with $C_{\rm spin}=0.8$ [Fig.~\ref{Fig_dtexture}~(b)] \cite{Comp_Spin}.

To simplify the analysis,
we write as $l_{\rm soliton} \equiv 4a+4r_{\rm core}$.
The equilibrium length is then determined by $\frac{\partial}{\partial l_{\rm soliton}}E_{\rm vortex}=\frac{\partial}{\partial l_{\rm soliton}}\left( U_{\rm hyd}+E_{\rm soliton} \right)=0$.
In the first approximation, the soliton energy $E_{\rm soliton}$ is expressed as $E_{\rm soliton}^{first}=\alpha_{\rm AF} l_{\rm soliton}$ with the tension coefficient $\alpha_{\rm AF}\sim \sqrt{q\mu} n_{\rm P}\xi_n$ of the AF-core soliton \cite{liu2020phase}.
This approximation fails for $\xi_q \gg \xi_s$.
Actually, the thickness of the elliptic core is much smaller than the thickness $\sim\xi_q$ of the AF-core soliton forming a halo structure [Fig.~\ref{Fig_dv}~(a)], which increases the tension effectively.
To take this effect into account, we introduce a phenomenological formula
\begin{eqnarray}
  \frac{E_{\rm soliton}}{\mu n_{\rm P}\xi_n^2} = \sqrt{\frac{q}{\mu}}\frac{l_{\rm soliton}}{\xi_n} \left( 1+ \frac{l_{\rm soliton}}{r_{\rm spin}} \right).
\label{eq:E_soliton}
\end{eqnarray}
This formula yields $\frac{l_{\rm soliton}}{\xi_n}=\frac{r_{\rm spin}}{4\xi_n}\left(\sqrt{ 1+ 8\pi\frac{\xi_q}{r_{\rm spin}}  } -1\right)$ and explains the scaling behavior $l_{\rm soliton}\sim l_{\rm spin}\propto q^{-0.25}$ for $\xi_q\gg \xi_s$ in Fig.~\ref{Fig_dtexture}(a).
This means that the soliton is effectively elastic with $E_{\rm soliton}\propto l_{\rm soliton}^2$ for $l_{\rm spin}\gg r_{\rm spin}$.

\begin{figure}
\begin{center}
\includegraphics[width=1.0 \linewidth,keepaspectratio]{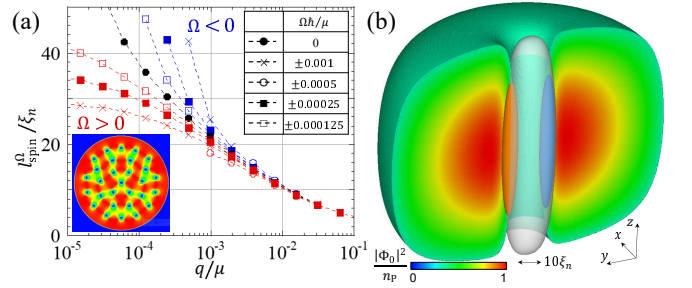}
\end{center}
\vspace{-5mm}
\caption{
(a) The $q$ dependence of the width $l_{\rm spin}^{\Omega}$ of a rotating elliptic vortex with angular velocity $\Omega$.
The single vortex is unstable for larger $|\Omega|$ and smaller $q$ due the boundary effect.
Inset: an elliptic-vortex lattice obtained after the instability due to the boundary effect for $\Omega\hbar/\mu=0.001$ and $q/\mu=2^{-11}$.
(b) The three-dimensional solution of an elliptic vortex in a harmonic trap for $\Omega \hbar/ \mu=0.0005$.
The isovolume plot shows the region $|\Psi_0|^2c_0/\mu \leq 0.3$ for $x>0$ with its $x=0$ cross-sectional profile.
A translucent surface along the $z$ axis  represents the isosurface of $|\Psi_{\pm 1}|^2c_0/\mu=0.15$,
 to which the two poles of the isosurfaces $s_yc_0/\mu=\pm 0.7$ are attached (red for positive and blue for negative).}
\label{Fig_rot}
\end{figure}

{\it Rotating solutions}.---Finally, the response to an external rotation is investigated as a dynamical property.
The external rotation of angular frequency $\Omega$ is described by the energy in the rotating frame $G'=G-\Omega L_z$, with the angular momentum $L_z$ along the $z$ axis \cite{landau1980statistical}.
The width $l_{\rm spin}^\Omega$ of an elliptic vortex decreases with $\Omega$ [Fig.~\ref{Fig_rot}(a)],
since the angular momentum increases more as the vorticity is localized more toward the center.
Owing to the boundary effect \cite{boundary_effect},
the single-vortex states are unstable for large $|\Omega|$ or small $q/\mu$, leading to a lattice of elliptic vortices [inset of Fig.~\ref{Fig_rot}(a)].

The three-dimensional structure of an elliptic vortex is demonstrated numerically for a feasible setup in Fig.~\ref{Fig_rot}(b).
A $^{23}$Na BEC of $5.6 \times 10^5$ atoms is in a harmonic trap $V_{\rm trap}=\frac{M}{2}(\omega_\bot^2r^2+\omega_z^2z^2)$ with $\frac{\hbar}{\mu}(\omega_\bot,\omega_z)\approx (0.019,0.024)$.
The spin spots appear as two poles (red and blue) along the $m=\pm 1$ component (translucent pole) in the vortex core. 

{\it Discussion}.---Although the wall-HQV composites were thought to be finally unstable, decaying into conventional axisymmetric vortices due to the snake instability of the wall \cite{kangPhysRevLett.122.095301,kangPhysRevA.101.023613},
 the result suggests that they survive as elliptic vortices after the phase transition.
The vortices including their dynamics will be observed through the transverse-spin spots by {\it in situ} magnetization imaging \cite{seoPhysRevLett.115.015301}.
 The theory here can be applied in a similar manner to the double-core vortex or the KLS-wall-HQV composite in $^3$He-B,
  while different forms of the hydrodynamic potential and soliton tension were introduced \cite{volovik1990half}.

  It is important to make a clear distinction between the types of HQVs in the AF phase (type I) and the P phase (type II).
  Properties of type-I HQVs are understood by the following correspondence between binary BECs and the AF phase.
  Since the equation of motion of spin-1 BECs with $\Phi_0=0$ reduces to that of binary BECs,
  HQVs in miscible binary BECs are physically identical to type-I HQVs in the absence of the $m=0$ component \cite{seoPhysRevLett.115.015301};
 type-I HQVs with the same circulation are repulsive according to Ref.~\cite{PhysRevA.83.063603},
  where the intra- and inter-component coupling constants correspond to $g_1=g_2=c_0+c_2$ and $g_{12}=c_0-c_2$, respectively.
  Therefore, a pair of type-I HQVs are unstable without external rotation \cite{PhysRevA.86.013613},
   which differs from type-II HQVs in that they form a bound pair by the wall tension \cite{Type-II,PhysRevLett.116.085301,PhysRevA.93.033633}.

 It should be mentioned that similar composite objects are investigated experimentally as the spin-mass vortex attached by a planar soliton in $^3$He-B \cite{kondoPhysRevLett.68.3331,Lounasmaa7760} and theoretically as the vortex molecules in Rabi-coupled binary BECs \cite{sonPhysRevA.65.063621,kasamatsuPhysRevLett.93.250406,ciprianiPhysRevLett.111.170401,tylutkiPhysRevA.93.043623,calderadoPhysRevA.95.023605,etoPhysRevA.97.023613,GallemiPhysRevA.100.023607,iharaPhysRevA.100.013630,kobayashiPhysRevLett.123.075303}.
Interestingly, the confinement of vortices by domain walls is considered a toy model of the quark-confinement problem \cite{sonPhysRevA.65.063621}.
 Accordingly, ``HQV-wall plasma'', an analog of quark-gluon plasma (QGP), occurs at a finite temperature $T$ at least for $T>\frac{q}{k_{\rm B}}$ in nematic-spin BECs, where thermal fluctuations free the spin spots from the confinement by the AF-core soliton.
In this sense, the observed phase-transition dynamics \cite{kangPhysRevLett.122.095301,kangPhysRevA.101.023613} are regarded as simulations of the transition dynamics from QGP to hadrons like the big bang simulation in `little bang' \cite{yagi2005quark}.
Further investigations on the dynamics and interactions of elliptic vortices will shed light on unexplored phase-transition dynamics in different physical systems.


\begin{acknowledgments}
H.T. thanks Yong-il Shin for discussion and critical reading of the Letter.
This work is supported by JSPS KAKENHI Grants No. JP17K05549, No. JP18KK0391, No. JP20H01842,
 and in part by the OCU ``Think globally, act locally'' Research Grant for Young Scientists through the hometown donation fund of Osaka City.
\end{acknowledgments}


\begin{thebibliography}{10}
\bibitem{Qc_vortex}
{ Neglecting the spin interaction with $\frac{c_1}{c_0}=0.016 \ll 1$, the
  critical value is determined $q_{\rm C}/\mu \approx 0.25$ by regarding
  $\Phi_{\pm 1}$ as a single-particle wave function bounded by the vortex core
  of $\Phi_0$. A similar problem is solved for the nematic-spin vortex in the
  AF phase \cite{underwood2020properties}. }.
\bibitem{boundary_effect}
{ In a rotating frame, the local superfluid velocity at the surface of the
  condensate is estimated as ${\bm v}_0=(v_\varphi-R\Omega) \hat{\bm \varphi}$
  with $v_\varphi \sim \frac{\hbar}{MR_\bot}$. When the velocity exceeds a
  critical value, comparable to the phonon velocity $\sqrt{\mu/M}$, the vortex
  nucleation occur after the excitation of ripplons due to the Landau
  instability at the surface. }.
\bibitem{Type-II}
{ Although type-II HQVs have been numerically realized under rotation
  \cite{PhysRevLett.116.085301,PhysRevA.93.033633}, they did not distinguish
  physically type-I and type-II by neglecting the impact of the domain wall. }.
\end{thebibliography}

\appendix

\renewcommand{\thesubsection}{A\arabic{subsection}}
\setcounter{subsection}{0}

\renewcommand{\theequation}{A\arabic{equation}}
\setcounter{equation}{0}
\renewcommand{\thefigure}{A\arabic{figure}}
\setcounter{figure}{0}

\section*{Supplemental material}

\subsection{Method of the numerical simulation}\label{ASec:numerical}

\begin{figure*}
\begin{center}
\includegraphics[width=1.0 \linewidth, keepaspectratio]{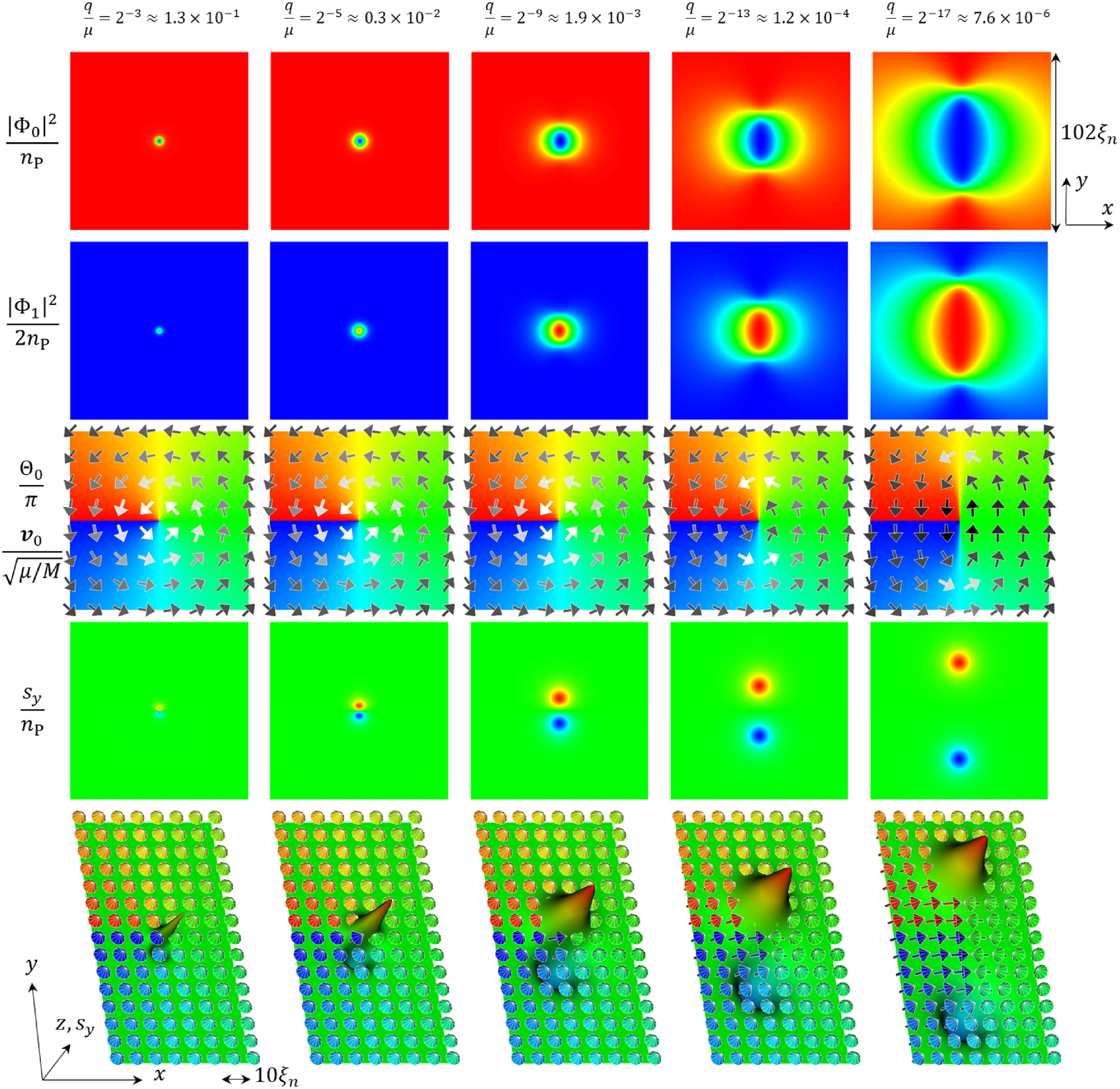}
\end{center}
\vspace{-5mm}
\caption{
The cross-sectional profiles of elliptic vortices and the three dimensional texture of pseudo-director field ${\bm g}/|{\bm g}|$ for $q=2^{-n_q}~(n_q=3,5,9,13,17)$.
The method of plot is the same as that of Fig.~\ref{Fig_dv} in the main text.
The angle of view is changed from that in Fig.~\ref{Fig_dv} in the three dimensional plots in the bottom.
}
\label{Fig_A1}
\end{figure*}

Here, we describe the method of numerical simulation used in this work.
The numerical solutions is obtained by minimizing the energy functional
$$
G''=\int^{R_1}_{-R_1} dx \int^{R_2}_{-R_2} dy \int^{R_3}_{-R_3} dz d^3x({\cal G}+V_{\rm trap}n-\Omega l_z)
$$
with the trapping potential $V_{\rm trap}$ and $l_z=\hbar\sum_m\Re[\Phi_m  x\partial_y\Phi_m-y\partial_x\Phi_m]$.
The space coordinates $(x,y,z)=(x_1,x_2,x_3)$ are discretized as $x_i(n_i)=-R_i+\Delta x n_i$ with $n_i=0,1,2,...N_i$ with $x_i(N_i)=R_i$.
The spatial derivatives of $\Phi_m$ are computed with finite difference approximation; e.g.,
 $\partial_x\Phi_m$ and $\partial_x^2\Phi_m$ are computed by the central difference of the first and second order, respectively.

 All solutions were obtained by minimizing the energy functional very carefully.
 The steepest descent method is performed by solving the imaginary time evolution $\frac{\partial \Phi_m}{\partial \tau}=-\frac{\delta G''}{\delta\Phi_m}$.
 The imaginary time $\tau$ is discretized as $\tau=\Delta \tau n_\tau$ with $n_\tau=0,1,2,....$.
 The time evolution is written as $\Phi_m(n_\tau+1)=\Phi_m(n_\tau)-\Delta\tau \frac{\delta G''}{\delta\Phi_m}(n_\tau)$.
 The evolutions were computed until the difference $G''(n_\tau)-G''(n_\tau-1000)$ becomes non-negative within the double precise by using Intel$\textsuperscript{\textregistered}$ Fortran Compiler.

The solutions in a uniform system is approximately obtained in a cylindrical box potential $V_{\rm trap}=V_0[\tanh(r-R)+1]$ with $V_0/\mu=20$, $R=0.95R_\bot$, and $R_1=R_2=R_\bot$.
Here,
we solve two-dimensional equations by assuming that the wave functions are homogeneous along the $z$ axis and thus independent of $z$.
The trap depth $V_0$ is taken to be so large that the order parameter damps quickly outside the cylinder and almost vanish nearby the system boundary.
The boundary effect becomes significant only when the distance $l_{\rm spin}$ between the spin spots becomes $\gtrsim R_\bot/2$.
The system size is set to be enough large to neglect the boundary effect for the results in the main text.
For the non-rotating case of $\Omega=0$ (the results of Figs.~\ref{Fig_dv} and \ref{Fig_dtexture}),
the numerical simulation was done with $2R_\bot=1024.5\xi_n$ with $N_1=N_2=2048$, $\Delta x=0.5\xi_n$, and $\Delta \tau =0.0025$.
It was confirmed that our results do not change essentially for $\Delta x=0.3\xi_n$ and $\Delta x=0.4\xi_n$ except for the finite-size effect,
 which is of no interest to our main subject.
 The finite-size effect becomes important only for $\frac{q}{\mu}\leq 2^{-19}\approx 1.9 \times 10^{-6}$ for $\Delta=0.5\xi_n$.
 For very small values of $\frac{q}{\mu}$ the width of the elliptic vortex become on the order of or larger than the system size and we could not obtain the vortex state.

The vortex solutions were obtained for $q/\mu=2^{-n_q}~(n_q=0,1,2,...)$ as shown in Fig.~\ref{Fig_A1}.
The vortex has the normal core with $\Phi_{\pm 1}=0$ for $n_q<3$ (not shown).
The protocol of the numerical simulation is as follows.
First, the solution for $n_q=3$ is obtained.
 Then, the initial state of the time evolution is set as $\Phi_0=f_0(r)e^{i\varphi}$ and $\Phi_{\pm 1} =f_{\pm 1}$ with $f_0(r)=\sqrt{\max(0,n'_{\rm TF})}$, $n'_{\rm TF}=c_0^{-1}(\mu-V_{\rm trap}-\frac{\hbar^2}{2Mr^2})$ and $f_{\pm 1}(r)=\pm \sqrt{\frac{n_{\rm P}}{2}}e^{-r^2/\xi_n^2}$.
The vortex can be stabilized in the center region even for the non-rotating case of $\Omega=0$ since the spatial gradient of $V_{\rm trap}$ is negligibly small there.
The solution for $n_q+1$ is obtained by using the solution of $n_q$ as the initial state.

The rotating case of Fig.~\ref{Fig_rot}~(a) is obtained with $2R_\bot=410\xi_n$ with $N_1=N_2=1048$ and $\Delta=0.4\xi_n$.
The protocol is the same as the non-rotating case.
 For the three dimensional simulation in the harmonic trap of Fig.~\ref{Fig_rot}~(b),
  the system size is $2R_\bot=192.5\xi_n$ and $2R_z=128.5\xi_n$ with $N_1=N_2=384$, $N_3=256$ and $\Delta=0.5\xi_n$.
  In the local density approximation, the effective chemical potential is written as $\mu'=\mu-V_{\rm trap}$.
  According to Fig. \ref{Fig_dtexture}(b), the spin density decreases with $q/\mu'$.
  This is why the spin poles becomes thinner as they are away from the trap center.
  The vortex core size becomes thicker as the local healing length $\frac{\hbar}{\sqrt{M\mu'}}$ becomes larger for large $|z|$.

  \subsection{Computation of the hydrodynamic potential}\label{ASec:HydP}

The velocity field ${\bm v}=(u,v)^T$ in a two dimensional potential flow is represented as
$u=\partial_y \Psi=\partial_x \Phi$ and
$v=-\partial_x \Psi=\partial_y \Phi$
with the Stream function $\Psi$ and the velocity potential $\Phi$.
The complex velocity potential $W=\Phi+i\Psi$ of a point vortex with a circulation $\Gamma$ in the complex plane $(x,y)$ is written as
$
W=-i\frac{\Gamma}{2\pi} \log~z
$.
The Joukowski transformation
$
z=\zeta+\frac{a^2}{\zeta}
$
with $\zeta=a e^{\phi +\psi}$
reads
$x=2a\cosh \phi \cos \psi$ and $y=2a \sinh \phi \sin \psi$.
This transformation corresponds to a mapping from a circle of radius $a$ to an ellipse of major radius $2a\cosh \phi$ and minor radius $2a \sinh\phi$ ($\phi \geq 0$) in the $xy$ plane.
The ellipse reduces a segment of length $4a$ along the $x$ axis for $\phi=0$.
The segment is along the $y$ axis if $a$ is replaced by $ia$ in the formula of $\zeta$.
We used the formula $\zeta=a e^{\phi +i\psi}$ in the following computation without loss of generality.

In a quantized vortex in a scalar superfluid, the velocity field diverges at the center of the vortex core, where the order parameter amplitude vanishes at the core.
The density increases to the bulk value far from the core.
The region within a circle of a radius $\xi_n$ with small density around the center is called the core region.
To evaluate the energy of a quantized vortex per unit length,
the contributions from the core region ($r<\xi_n$) and its outer region ($r>\xi_n$) is computed separately.
Similarly, for the elliptic vortex, there exists the core region of an elliptic form around the band-shaped singularity and the energy is computed separately.

To compute the energy analytically, we neglect the so-called quantum pressure term in the Thomas-Fermi (TF) approximation \cite{pethick2008bose}.
Then, the density far from the vortex core can be written as
$$
n \approx n_{\rm TF}=n_{\rm bulk}\left(1-\frac{M}{2\mu}{\bm v}^2\right)
$$
with $n_{\rm bulk}=n_{\rm P}$ is the bulk density.
In this approximation, one obtains the contribution to the energy functional $G$ from the outer region of area $S_{\rm out}$ up to the order of ${\cal O}\left(\frac{M}{2\mu}{\bm v}^2\right)$,
\begin{eqnarray}
E_{\rm out}
&=&\int_{S_{\rm out}}d^2x {\cal G}
\nonumber \\
&\approx&\int_{S_{\rm out}}d^2x\left[\frac{Mn_{\rm TF}}{2}{\bm v}^{2} +{\cal U} (n_{TF})\right]
\nonumber \\
&=& \int_{S_{\rm out}}d^2x \left[\frac{Mn_{\rm P}}{2}{\bm v}^{2} +{\cal U}_{\rm P}\right].
\end{eqnarray}
Here, ${\cal U} (n_{TF})$ is the energy density ${\cal U}$ evaluated in the TF approximation, and it reduces to, for the bulk P phase,
$$
{\cal U}_{\rm bulk}={\cal U}_{\rm P}=-\frac{\mu^2}{2c_0}.
$$

A local state, different from the P state, appears in the core region where the $m=0$ component vanishes.
The contribution from the core region is written as
$$
E_{\rm core}={\cal U}_{\rm core}S_{\rm core}
$$
with the energy density ${\cal U}_{\rm core}$ and the area $S_{\rm core}$ of the core region.

The vortex energy $E_{\rm vortex}$ is defined as an excess energy in the presence of the vortex,
the difference between the total energy with a vortex and the energy $E_{\rm bulk}={\cal U}_{\rm bulk}(S_{\rm core} +S_{\rm out})$ in the absence of it;
\begin{eqnarray}
E_{\rm vortex}
&=& E_{\rm out}+E_{\rm core}-E_{\rm bulk}
\nonumber \\
&=& U_{\rm out}+U_{\rm core}
\end{eqnarray}
with $U_{\rm core(out)}=E_{\rm core(out)}-{\cal U}_{\rm P}S_{\rm core(out)}$.
The potentials of the outer and core regions are rewritten as
\begin{eqnarray}
U_{\rm out} &=& \frac{Mn_{\rm bulk}}{2}\int _{S_{\rm out}} d^2x{\bm v}^{2}
\nonumber \\
U_{\rm core} &=& \delta \mu n_{\rm bulk} S_{\rm core}
\end{eqnarray}
with $\displaystyle \delta =\frac{{\cal U}_{\mathrm{core}}}{\mu n_{\mathrm{bulk}}} +\frac{1}{2}$.

The potential $U_{\rm out}$, which is reduced to the hydrodynamic potential $U_{\rm hyd}$ as shown later, is evaluated by computing the integral
\begin{eqnarray}
I&=&
\int _{S_{\rm out}} dxdy {\bm v}^{2}
\nonumber \\
&=&
\left(\frac{\Gamma }{2\pi }\right)^{2}\int _{S_{\rm out}} dxdy\left| \frac{1}{\sqrt{z^{2} -4a^{2}}}\right| ^{2}
\nonumber\\
&=&
\left(\frac{\Gamma }{2\pi }\right)^{2}\int _{S_{\rm out}} dxdy\frac{1}{\sqrt{\left( x^{2} -y^{2} -4a^{2}\right)^{2} +4x^{2} y^{2}}}.
\nonumber
\end{eqnarray}
Here, we used
$$
|u|= \frac{\Gamma }{2\pi } \left|\Im \left[\frac{\partial _{x} \zeta }{\zeta }\right] \right| = \frac{\Gamma }{2\pi }\left| \Im \left[\frac{1}{\sqrt{z^{2} -4a^{2}}}\right] \right|
$$
and
$$
|v|=\frac{\Gamma }{2\pi }\left| \Im \left[\frac{\partial _{y} \zeta }{\zeta }\right] \right|=\pm \frac{\Gamma }{2\pi }\left| \Re \left[\frac{1}{\sqrt{z^{2} -4a^{2}}}\right]\right|.
$$

According to the transformation (for the ellipse along the $x$ axis)
$$
(x,y)=(2a\cosh \phi \cos \psi ,2a\ \sinh \phi \sin \psi )
$$
we have the determinant of the Jacobian matrix
$$
\left| \frac{\partial ( x,y)}{\partial ( \phi ,\psi )}\right|
= 2a^{2}(\cosh 2\phi -\cos 2\psi ).
$$
The integral $I$ is computed as
\begin{eqnarray}
  I
  &=&
  \left(\frac{\Gamma }{2\pi }\right)^{2}\int ^{\phi _{R}}_{\phi _{\rm core}} d\phi \int ^{\pi }_{-\pi } d\psi \frac{1}{2a^{2}\sqrt{(\cosh 2\phi -\cos 2\psi )^{2}}}\left| \frac{\partial ( x,y)}{\partial ( \phi ,\psi )}\right|
  \nonumber \\
  &=&
  \frac{\Gamma ^{2}}{2\pi }\int ^{\phi _{R}}_{\phi _{\rm core}} d\phi
  \nonumber\\
 &=&\frac{\Gamma ^{2}}{2\pi }( \phi _{R} -\phi _{\rm core})
 \nonumber\\
 &=&\frac{\Gamma ^{2}}{2\pi }\ln\frac{\rho _{R}}{\rho _{\rm core}}
 \nonumber
\end{eqnarray}
with
the radius of the system boundary in the $\zeta$ plane
$$
\rho _{R} =ae^{\phi _{R}} ~{\rm or}~ \phi _{R} =\ln\frac{\rho _{R}}{a}
$$
and the cutoff radius for the core region
$$
\rho _{\rm core} =ae^{\phi _{\rm core}} ~{\rm or}~ \phi _{\rm core} =\ln\frac{\rho _{\rm core}}{a} \ ( \rho _{\rm core}  >a).
$$
The system boundary and the cutoff circle in the $\zeta$ plane are mapped into ellipses in the $xy$ plane.
The major and minor radiuses are written as
\begin{eqnarray}
&&
A_{R,{\rm core}} =2a\cosh \phi _{R,{\rm core}} =\frac{\rho ^{2}_{R,{\rm core}} +a^{2}}{\rho _{R,{\rm core}}}
\nonumber\\
&& B_{R,{\rm core}} =2a\sinh \phi _{R,{\rm core}} =\frac{\rho ^{2}_{R,{\rm core}} -a^{2}}{\rho _{R,{\rm core}}}
\end{eqnarray}
and they satisfy the relation
$$
 A_{R,{\rm core}} +B_{R,{\rm core}} =2\rho _{R,{\rm core}}.
$$
Here, $A_{\rm core}$ and $B_{\rm core}$ correspond to $R_{+}$ and $R_{-}$ in the main text, respectively.
In the limit $\frac{\rho_R}{a} \to \infty$, we have
$$
A_R=B_R \to \rho_R =R
$$
with the radius $R$ of the system boundary in the $xy$ plane.

The size of the core region is parametrized by two parameters, $a$ and
$$
r_{\rm core}\equiv \rho_{\rm core}-a.
$$
Then, we have
$$
U_{\rm out}\approx U_{\rm hyd}=\frac{Mn_{\rm P}}{2}\frac{\Gamma^2}{2\pi}\ln \frac{R}{a+r_{\rm core}}.
$$
The area $S_{\rm core}$ of the core region is represented as
$$
S_{\rm core}=\pi A_{\rm core} B_{\rm core} =\pi \frac{( a+r_{\rm core})^{4} -a^{4}}{( a+r_{\rm core})^{2}}
$$
The oblateness $f$ of the core region is defined as
$$
f=1-\frac{B_{\rm core}}{A_{\rm core}} =\frac{2}{1+( 1+r_{\rm core} /a)^{2}}.
$$
For $\frac{r_{\rm core}}{a}\ll1$, asyptotic to the limit of the maximum oblateness ($f=1$),
we have
$$
A_{\rm core} \approx 2a,~~B_{\rm core} \approx 2 r_{\rm core}
$$
and for $\frac{r_{\rm core}}{a}\gg 1$, asyptotic to the limit of the minimum oblateness ($f=0$),
$$
A_{\rm core} = B_{\rm core} \approx r_{\rm core}.
$$
The former limit corresponds to a vortex with a band-shaped core region in three dimensions, called a vortex band.
The latter corresponds to a conventional vortex filament with cylindrical core region.

\subsection{Computation of the spin interaction}\label{ASec:Espin}
The local BA state, emerging around the edge of the elliptic vortex, is collateral in the existence of the local AF state in the vortex center.
Therefore, the magnetization can be associated with the density at the origin $n_{\rm core}=n({\bm r}=0)$.
According to the mean-field approach in the previous studies \cite{liu2020phase,underwood2020properties},
 a continuous phase transition occurs at a critical point ($q=q_{\rm C}$) in the core of a topological defect.
This approach is also well applicable to our case.
We obtain similar behaviors $n_{\rm core} \propto 1-\frac{q}{q_{\rm C}}$ and $s_y^{\rm max}=\max s_y\propto \sqrt{1-\frac{q}{q_{\rm C}}}$.
Since the density is asymptotic to $n_{\rm AF}$ far from the critical point for $q \ll q_{\rm C}$,
 a quantitative estimation is obtained by
$$
n_{\rm core}\sim n_{\rm AF}\left(1-\frac{q}{q_{\rm C}}\right),
$$
which is quantitatively agreed with the numerical result [Fig.~\ref{Fig_dtexture}(b)].

The magnetization is well described with this approach too.
The magnetization happens in the two spin spots with $\arg \Phi_0=\pm \pi$ along the $y$ axis.
Then the local spin density is written as $F_y\sim \pm 2\sqrt{2n_1 |\Phi_0|^2}$ with $\Phi_{+1}=-\Phi_{-1}=\sqrt{n_1}\propto\sqrt{n_{\rm core}}$.
The density is almost constant $n\approx n_{\rm P}$ everywhere for small $q$ and then the maximum value is estimated by the relation between the arithmetic and geometric means with $2n_1=\frac{n}{2}\approx \frac{n_{\rm core}}{2}$ and $|\Phi_0|^2=\frac{n}{2}\approx \frac{n_{\rm P}}{2}$ as
$$
s_y^{\rm max} \sim \frac{\sqrt{n_{\rm core}n_{\rm P}}}{1+\frac{c_2}{c_0}},
$$
where the factor in the denominator comes from the spin interaction in the presence of spin density.

The size $r_{\rm spin}$ of the magnetic spot around the edges grows with the size of the AF-core, $\sim\xi_q$.
The spin interaction becomes more important as the magnetic spot grows and the spot size finally reaches the spin healing length,
estimated as
$$
\xi_s =\frac{\hbar}{\sqrt{M\sigma}}
$$
with $\sigma= c_2n_{\rm P}$.
This crossover behavior of the spot size is described by a simple formula
$$
r_{\rm spin}=\frac{1}{\xi_s^{-1}+C_{\rm spin}\xi_q^{-1}},
$$
with a constant $C_{\rm spin}\sim {\cal O}(1)$.
Finally, the spin interaction energy is evaluated as
$$
E_{\rm spin}\sim \frac{1}{2}c_2(s_y^{\rm max})^2\pi r_{\rm spin}^2
\sim \frac{\pi}{2}\frac{c_2}{c_0}\frac{n_{\rm core}}{n_{\rm P}}\left(\frac{r_{\rm spin}}{\xi_n}\right)^2\mu n_{\rm P}\xi_n^2.
$$
This formula is well-consistent with the numerical result with $C_{\rm spin}=0.8$.


\end{document}